\documentclass[letterpaper,peerreviewca,12pt]{IEEEtran}
\usepackage{amsmath, amssymb}
\usepackage{graphicx,wrapfig}

\newtheorem{theorem}{Theorem}
\newtheorem{lemma}{Lemma}

\newcommand{\NP}{{\textsf{NP}}}
\newcommand{\coNP}{{\textsf{coNP}}}
\newcommand{\PSPACE}{{\textsf{PSPACE}}}

\begin{document}

\title{Solving Classical String Problems\\ on Compressed Texts}
%\\[Extended Abstract]

\author{
\authorblockN{Yury Lifshits\thanks{Support
by grant INTAS  04-77-7173 and contract with Federal Agency of Science and Innovation ¹02.442.11.7291 are gratefully acknowledged.} }
\authorblockA{Steklov Institute of Mathematics\\
St.Petersburg, Russia\\
Email: yura@logic.pdmi.ras.ru} }

%\specialpapernotice{[Extended Abstract]}

\maketitle

 \begin{abstract}
Here we study the complexity of string problems as a function of {\it the size of a program that
generates input}. We consider {\it straight-line programs} (SLP), since all algorithms on SLP-generated
strings could be applied to processing LZ-compressed texts.

The main result is a new algorithm for pattern matching when both a text $T$ and a
pattern $P$ are presented by SLPs (so-called fully compressed
pattern matching problem). We show how to find a first occurrence, count all
occurrences, check whether any given position is an occurrence or not in time $O(n^2m)$.
Here $m,n$ are the sizes of straight-line programs generating correspondingly
$P$ and $T$. 

Then we present polynomial algorithms for computing fingerprint table and
compressed representation of all covers (for the first time) and for
finding periods of a given compressed string (our algorithm is faster
than previously known). On the other hand, we show that computing
the Hamming distance between two SLP-generated strings is \NP- and \coNP-hard. 
\end{abstract}
 
%\keywords{Pattern matching, algorithm, compressed text.}

%\IEEEpeerreviewmaketitle

\section{Introduction}

{\bf Background.} How to solve text problems, if instead of an input
string we get only program generating it? Is it possible to
find a faster solution than just ``generate text + apply classical
algorithm''? In this paper we consider strings generated by {\it straight-line
programs} (SLP). These are programs using only assignment operator.
The exact definition and discussion on this notion are given in
Section~\ref{slp}.

We come to this question studying algorithms on compressed texts. Many algorithms for direct
search in compressed texts without unpacking were proposed in the last decade. Finally
(see Rytter's work \cite{Ry03}), it turns out that only decompression stage really matters.
More precisely, we can consider a pair (decompression algorithm, archive file) as a {\it generating description}
for the original text. Rytter presents an effective translation from any LZ-compressed text to the straight-line
program generating the same text.

Here we study complexity of solving problems as a function of {\it the size of a program that
generates input}. Since the ratio between the size of SLP and the length of original string
might be exponential, even any polynomial algorithm is a matter of interest.
Main purpose of such algorithms is to speed up naive approach and to save memory
(since we never generate the original text). Moreover, even if we have just original text data,
but containing a lot of repetitions, we can find generating description first and
then apply one of our algorithms for SLP-generated texts. A similar
idea (for OBDD-described functions) is used in symbolic model checking.

There are already promising applications for work with compressed
objects: solving word equations in polynomial space \cite{Pl04},
and pattern matching in message sequence charts \cite{GeMu02}.
Potential applications include all fields with highly
compressible data like bio-informatics (genomes contain a lot of
repetitions), pattern matching in statistical data (like internet
log files) and any kinds of automatically generated texts.

%in models of parallel systems (straightforward
%description usually require exponential number of states).

{\bf Problem.} For any text problem on SLP-generated strings we ask two questions: (1) does any
polynomial algorithm exist? (2) If yes, what is exact complexity of the problem? We can think about
negative answer to the first question (say, NP-hardness) as an evidence that naive ``generate-and-solve''
is the best way for that particular problem.    

First of all we consider complexity of pattern matching problem on SLP-generated strings
(sometimes called fully compressed pattern matching or FCPM). 
That is, given a SLPs generating pattern $P$ and text $T$ to
answer whether $P$ is a substring of $T$ and provide a succinct description of all
occurrences. An important special case is {\it equivalence problem} for SLP-generated texts.
Then we expand the obtained algorithm/technique for several other string problems: finding shortest period/cover and 
constructing fingerprint table. Finally we consider the Hamming distance problem. 

{\bf Results.} The key result of the paper is a new $O(n^2m)$ algorithm for pattern matching 
on SLP-generated texts, where $m$ and $n$ are sizes of SLPs generating $P$ and $T$,
correspondingly. This is an improvement of the
$O((n+m)^5\log^3|T|)$ algorithm by G\c{a}sieniec et. al
\cite{GaKaPlRy96} and $O(n^2m^2)$ algorithm by Miyazaki et al.
\cite{MiShTa97}. For one quite special class of SLP the FCPM problem was
solved in time $O(mn)$ \cite{HiShTaAr00}.

After presenting the main algorithm in Section~\ref{algo}
we sketch (more details will follow in a full version of paper)
the polynomial algorithms for finding covers and periods, and for computing a fingerprint table
for SLP-generated texts in Section~\ref{other}. On the other hand, we show that surprisingly,
computing the Hamming distance is \NP-hard. Here we have a closely
related problems (Hamming distance and equivalence) from different
sides of the border between efficiently solvable problems on
SLP-generated texts and intractable ones. 
%Hence, there is some difference i

{\bf Proofs.} The main ingredients of the algorithm are
dynamic programming method, operations with arithmetical
progressions and special tricks in case of dense sequence of
pattern occurrences. The following way to work with SLP-generated strings
turns out to be the most productive: (1) invent some auxiliary property of strings
(2) compute it for all intermediate texts (3) derive the answer from the computed array.
In our pattern matching algorithm we consecutively compute elements of a
special $n\times m$ table. This is exactly the same idea as
in \cite{MiShTa97}. However, the routine for computing a new element is completely different: we
use only $O(n)$ time while in the paper \cite{MiShTa97} $O(mn)$
time is used. One of the key tricks comes from checking occurrence
of false mismatches for Rabin-Karp algorithm as in \cite{Ma93}.
However we just apply the same intuition but not directly the same
procedure.

Immediate consequence of our result is an $O(n^2m)$ algorithm for pattern matching in
LZ-compressed texts. Our algorithm uses only linear time for one step of dynamic programming. Hence,
in order to get faster method, we now need a radically new approach.

\subsection{Related results}

The whole field started from papers by Amir, Benson and Farach
\cite{AmBeFa94} and by Farach and Thorup \cite{FaTh95} presenting
algorithms for compressed pattern matching with working time
depending on {\it size of compressed text}. Compression models
vary from run-length encoding, different members of Lempel-Ziv
family \cite{Zile77}, straight-line programs (SLP) and collage
systems \cite{KiMaShTaShAr03}. The last two are good theoretical
models that generalize all previously studied compression
algorithms. Namely, Rytter \cite{Ry03} proved that any LZ-encoding
could be efficiently translated to SLP of approximately the same
size without unpacking.

In addition to pattern matching, also algorithms for window
subsequence (i.e. scattered substring) search \cite{CiGuLiMa06},
membership in regular language \cite{Na03}, and approximate
pattern matching \cite{KaNaUk00} were presented. From the other
side, it was shown that fully compressed subsequence problem and
the longest common subsequence are $\Theta_2$-hard (this is a kind
of closure of \NP\ and \coNP), compressed and fully compressed
two-dimensional pattern matching are \NP-complete and
$\Sigma_2$-complete, respectively \cite{BeKaLaPlRy02}, while
membership in a context-free language is even \PSPACE-complete
\cite{Lo04}.

Encouraging experimental results are reported in \cite{NaRa99}.
The papers \cite{Ry00,Ry04} survey the field of processing
compressed texts, while the book \cite{Gu97} provides a general
overview of classical pattern matching algorithms.

\section{Compressed Strings Are Straight-Line Programs}\label{slp}

A {\it Straight-line program} is a context-free grammar generating
exactly one string. Moreover, we allow only two types of
productions: $X_i\rightarrow a$ and $X_i\rightarrow X_pX_q$ with
$i>p,q$. The string presented by a given SLP is a unique text
corresponding to the last nonterminal $X_m$. Although in previous
papers $X_i$ denotes only a nonterminal symbol while the
corresponding text was denoted by $val(X_i)$ or $eval(X_i)$ we
will identify this notions and use $X_i$ both as a nonterminal
symbol and as the corresponding text. We say that the size of SLP
is equal to the number of productions.

{\bf Example.} Consider string $abaababaabaab$. It could be generated by the following SLP: 
$$X_7\rightarrow X_6X_5, \quad X_6\rightarrow X_5X_4, \quad X_5\rightarrow X_4X_3, \quad X_4\rightarrow X_3X_2, 
\quad X_3\rightarrow X_2X_1, \quad X_2\rightarrow a, \quad X_1\rightarrow b.$$

\begin{wrapfigure}[10]{r}{5.75cm}
\includegraphics[scale=.5]{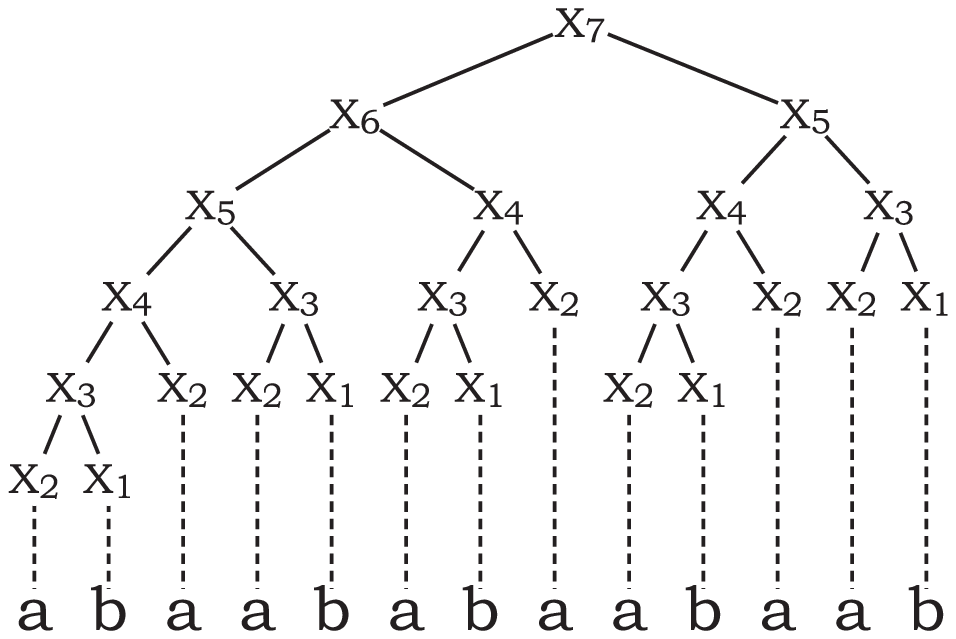}
\end{wrapfigure}

In fact, the notion of SLP describes only {decompression}
operation. We do not care how such an SLP was obtained.
Surprisingly, while the compression methods vary in many practical
algorithms of Lempel-Ziv family and run-length encoding, the
decompression goes in almost the same way. In 2003 Rytter
\cite{Ry03} showed that given a LZ-encoding of string $T$ we
could efficiently get an SLP encoding for the same string which is
at most $O(\log|T|)$ times longer than the original LZ-encoding.
This translation allows us to construct algorithms only in the
simplest SLP model. If we get a different encoding, we just
translate it to SLP before applying our algorithm. Moreover, if
we apply Rytter's translation to LZ77-encoding of a string $T$,
then we get an $O(log|T|)$-approximation of the {\it minimal} SLP generating
$T$. 

The straight-line programs allow the exponential ratio between
the size of SLP and the length of original text. Consider, for example
$$X_n\rightarrow X_{n-1}X_{n-1}, \qquad \dots \qquad
X_2\rightarrow X_1X_1, \qquad X_1\rightarrow a.$$

In complexity analysis we use both $\log |T|$ and $n$ (number of rules in SLP).
For example, we prefer $O(n\log|T|)$ bounds to $O(n^2)$, since in practice
the ratio between the size of SLP and the length of the text might be much
smaller than exponential.
%$\log|T|$

\section{A New Algorithm for Fully Compressed Pattern Matching}\label{algo}

Decision version of the fully compressed pattern matching problem (FCPM) is as follows:

\begin{center}
\fbox{\begin{minipage}{10cm}
 {INPUT:} Two straight-line programs generating $P$ and $T$
 \par {OUTPUT:} Yes/No (whether $P$ is a substring in $T$?)
\end{minipage}}
\end{center}

Other variations are: to find the first occurrence, to count all
occurrences, to check whether there is an occurrence from the given
position and to compute a ``compressed'' representation of all
occurrences. Our plan is to solve the last one, that is, to
compute auxiliary data structure that will contain all necessary
information for effective answering on the other questions.

We need some notation and terminology. We call a {\it position} in the text
a point between two consequent letters. Hence, the text $a_1\dots a_n$ has positions
$0,\dots,n$ where first is in front of the first letter and the last one after the last letter.
We say that some substring {\it touches} a given position if this position is either inside or on the border
of this substring. We use the term {\it occurrence} both as a corresponding substring and as its starting position.
Hopefully, the right meaning is always clear from the context.

Let $P_1, \dots, P_m$ and $T_1, \dots,T_n$ be the nonterminal
symbols of SLPs generating $P$ and $T$. For each of these texts we
define a special {\it cut} position. It is a starting position for
one-letter texts and merging position for $X_i=X_rX_s$. In the example above, 
the cut position for the intermediate text $X_6$ is between 5th and 6th letters: $abaab|aba$,
since $X_6$ was obtained by concatenating $X_5=abaab$ and $X_4=aba$.

We use a computational assumption which was used in all previous
algorithms but not stated explicitly. In analysis of our algorithm
we count arithmetical operations on text positions as unit
operations. In fact, text positions are integers with at most
$\log|T|$ bits in binary form. Hence, the bit operation complexity
is larger than our $O(n^2m)$ up to some $\log|T|$-dependent
factor.

Explanation of the algorithm goes in three steps. We introduce a
special data structure ({\it AP-table}) and show how to solve
pattern matching problem using this table in
Subsection~\ref{idea}. Then we show how to compute AP-table using
{\it Local PM} procedure in Subsection~\ref{apt}. Finally, we
present an algorithm for Local PM in Subsection~\ref{localpm}.

\subsection{Idea of the algorithm}\label{idea}

Our algorithm is based on the following theoretical fact (it was already used in \cite{MiShTa97}):

\begin{lemma}[Basic lemma]
All occurrences of $P$ in $T$ touching any given position
form a single arithmetical progression (ar.pr.)
\end{lemma}

\begin{center}
\includegraphics{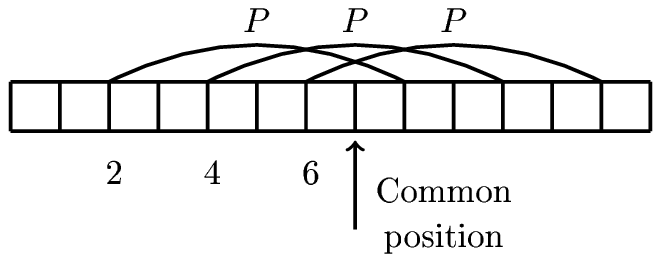}
\end{center}

The {\it AP-table} is defined as follows. For every $1\leq i\leq m, 1\leq j\leq n$
the value $AP[i,j]$ is a code of ar.pr. of occurrences of $P_i$ in $T_j$
that touch the cut of $T_j$. Note that any ar.pr. could be encoded by three integers:
first position, difference, number of elements. If $|T_j|<|P_i|$ we define $AP[i,j]=0$,
and if text is large enough but there are no occurrences we define $AP[i,j]=\varnothing$

\begin{center}
\includegraphics{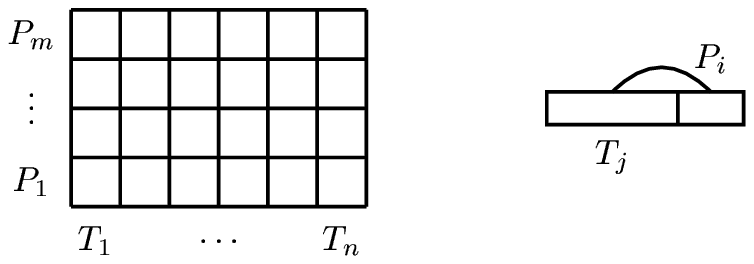}
\end{center}

\textbf{Claim 1:} Using AP-table (actually, only top row is
necessary) we can solve decision, count and checking versions of
FCPM in time $O(n)$.
\par \textbf{Claim 2:} We can count the whole AP-table by dynamic programming method in time $O(n^2m)$.

\medskip
{\bf Proof of Claim 1.} We get the answer for decision FCPM by the
following rule: $P$ occurs in $T$ iff there is $j$ such that
$AP[m,j]$ is nonempty. Checking and counting are slightly more
tricky. Recursive algorithm for checking: test whether the
candidate occurrence touches the cut in the current text. If yes,
use AP-table and check the membership in the corresponding ar.pr.,
otherwise call recursively this procedure either for the left or
for the right part. We will inductively count the number of
$P$-occurrences in all $T_1,\dots,T_n$. To start, we just get the
cardinality of the corresponding ar.pr. from AP-table. Inductive
step: add results for the left part, for the right part and
cardinality of central ar.pr. without ``just-touching''
occurrences.

\subsection{Computing AP-table}\label{apt}

\begin{center}
\fbox{\begin{minipage}{15cm}
 Sketch of the algorithm for computing AP-table. 
\begin{enumerate}
	\item Precomputation: compute lengths and cut positions for all intermediate texts; 
	\item Compute the first row and the first column of AP-table;
	\item From j=2 to n do  From i=2 to m do Compute AP[i,j]
	      \begin{enumerate}
	      \item Compute occurrences of the larger part of $P_i$ in $T_j$ around the cut of $T_j$;
        \item Compute occurrences of the smaller part of $P_i$ in $T_j$ that starts at ending positions of the larger                 part occurrences;
        \item Intersect occurrences of smaller and larger part of $P_i$ and merge all results to a single ar.pr.
        \end{enumerate}
\end{enumerate}
\end{minipage}}
\end{center}

At the very beginning we inductively compute tables for lengths, cut
positions, first letter, last letter of the texts
$P_1,\dots,P_m,T_1,\dots,T_n$ in time $O(n+m)$.

We compute elements of AP-table one-by-one. At the beginning we compute the first row and the first
column, then in the following order:
 \begin{center} From j=2 to n do  From i=2 to m do Compute AP[i,j] \end{center}

Case of $|P_i|=1$: compare with $T_j$ if $|T_j|=1$ or compare with
the last letter of the left part and the first one of the right part (we get this letters from precomputation stage). The resulting ar.pr. will be one or two neighbor
positions. Hence, just O(1) time used for computing every cell in the table.

Case of $|T_j|=1$: if $|P_i|>1$ return $0$, else compare letters. Also $O(1)$ time is enough for
every element.

Induction step: let $P_i$ and $T_j$ be both of length greater than
one. We are going to compute a new element in time $O(n)$ using
already computed values of AP-table. For this purpose, we design a
special auxiliary procedure that extracts useful information from
already computed part of AP-table.

This is a border between ideas of Miyazaki et al. and the new approach presented in the paper.
They compute the same table, but new element routines are completely different.

Procedure $Local PM(i,j,[\alpha,\beta])$ returns
occurrences of $P_i$ in $T_j$ inside the interval $[\alpha,\beta]$.

Important properties:
\begin{itemize}
 \item Local PM uses values AP[i,k] for $1\leq k\leq j$,
 \item It is defined only when $|\beta-\alpha|\leq 3|P_i|$,
 \item It works in time $O(j)$,
 \item The output of Local PM is a pair of ar.pr., all occurrences inside each ar.pr. have a common position,
 and all elements of the second are to the right of all elements of the first.
\end{itemize}

We now show how to compute a new element using 5 Local PM calls.
Let $P_i=P_rP_s$, the cut position in $T_j$ be $\gamma$ (we get it from precomputation stage) and,
without loss of generality, let $|P_r|\geq|P_s|$. The intuitive
way is (1) to compute all occurrences of $P_r$ ``around'' cut of
$T_j$, (2) to compute all occurrences of $P_s$ ``around'' cut of
$T_j$, and (3) shift the latter by $|P_r|$ and intersect.

\begin{center}
\includegraphics{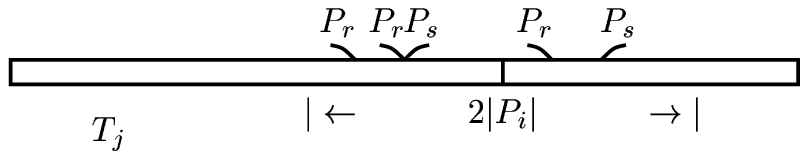}
\end{center}

Any occurrence of $P_i$ that touches the cut $\gamma$ of $T_j$ consists
of occurrence of $P_r$ and occurrence of $P_s$ inside $|P_i|$-neighborhood
of cut where ending position of $P_r$ equals starting position of $P_s$.

We apply Local PM for finding all occurrences of $P_r$ in the
interval $[\gamma-|P_i|,\gamma+|P_r|]$. Its length is
$|P_i|+|P_r|\leq 3|P_r|$. As an answer we get two ar.pr. of all
potential starts of $P_i$ occurrences that touch the cut.
Unfortunately, we are not able  to do the same for $P_s$, since
the length of interesting interval is not necessarily constant in
terms of $|P_s|$. %%[Explain more!]
So we are going to find only occurrences of $P_s$ that are
starting from two arithmetical progressions of endings of $P_r$
occurrences.

We will process each ar.pr. separately. We call an ending {\it continental} if
it is at least $|P_s|$ far from the last ending in progression, otherwise we call
it {\it seaside}.

\begin{center}
\includegraphics{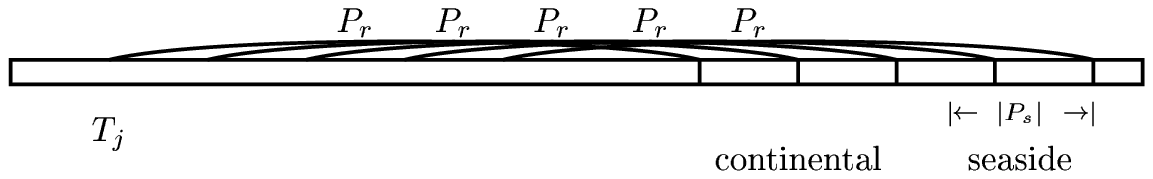}
\end{center}

Since we have an ar.pr. of $P_r$ occurrences that have common
position (property 4 of Local PM), all substrings of length $|P_s|$
starting from continental endings are identical. Hence, we need to
check all seaside endings and {\it only one} continental position.
For checking the seaside region we just apply Local PM for
$|P_s|$-neighborhood of last endpoint and intersect the answer
with ar.pr. of seaside ending positions. Intersecting two ar.pr.
could be done in time $O(\log|T|)$ by technique similar to
extended Euclid algorithm. For checking continental region we
apply Local PM for $|P_s|$ substring starting from the first
continental ending.

As the answer we obtain all continental endings/or none of them,
plus some sub-progression of seaside endings, plus something
similar for the second ar.pr. Since all of these four parts are
ar.pr. going one after another, we could simplify the answer to one
ar.pr. (it must be one ar.pr. by Basic Lemma) in time $O(1)$.

Let us estimate the complexity of the procedure for computing a new element. We use one Local PM call for $P_r$,
four Local PM calls for $P_s$, and twice compute the intersection of arithmetical progressions. Hence, we
have 7 steps of $O(n)$ complexity.

\subsection{Realization of Local PM}\label{localpm}

By the first step (crawling procedure) we put all occurrences of $P_i$ in $[\alpha,\beta]$ interval of $T_j$ to the
list. On the second step (merging procedure) we merge all these occurrences to just two ar.pr.

In crawling procedure on $(i,j,[\alpha,\beta])$ we take the ar.pr.
of occurrences of $P_i$ in $T_j$ that touch the cut, leave only
occurrences within the interval, and output this truncated ar.pr.
to the list. After that, we check whether the intersection of the
interval $[\alpha,\beta]$ with left/right part of $T_j$ is at
least $|P_i|$ long. If so, we recursively call crawling procedure
with the same $i$, the index of left/right part of $T_j$, and with
this intersection interval.

Consider the set of all intervals we work with during the crawling
procedure. Note that, by construction, any pair of them are either
disjoint or embedded. Moreover, since the initial interval was at
most $3|P_i|$, there are no four pairwise disjoint intervals in
this set. If we consider a sequence of embedded intervals, then
all intervals correspond to their own intermediate text from
$T_1,\dots,T_j$. Therefore, there were at most $3j$ recursive
calls in crawling procedure and it works in time $O(j)$.

We assume that crawling procedure always maintain the pointer to the corresponding place in the list. This
means that at the end we get a sorted list of at most $3n$ arithmetical progressions. By ``sorted''
we mean that the last element of $k$-th progression is less than or equal to the first one of $k+1$-th progression.
It follows from construction of crawling procedure, that output progressions could have only first/last
elements in common.

Now in the merging procedure we go through the resulting list of
progressions. Namely, we compare the distance between the last
element of current progression and the first element of the next
progression with the differences of these two progressions. If all
three numbers are equal we merge the next progression with the
current one. Otherwise we just move to the next progression.

If we apply the Basic Lemma to $\delta_1=\frac{2\alpha+\beta}{3}$
and $\delta_2=\frac{\alpha+2\beta}{3}$ positions, we will see that
all occurrences of $P_i$ in $[\alpha,\beta]$ interval form two
(one after another) arithmetical progressions. Namely, those who
touch $\delta_1$ and those who don't touch but touch $\delta_2$. Here we
use that $\beta-\alpha\leq 3|P_i|$, and therefore any occurrence
of $P_i$ touches either $\delta_1$ or $\delta_2$. Hence, our
merging procedure will start a new progression at most once.

\subsection{Discussion on the Algorithm}\label{discussion}

Here we point out two possible improvements of the algorithm.
Consider in details the ``new element routine''. Note that Local
PM uses only $O(h)$ time, where $h$ is the height of the SLP
generating $T$, while intersection of arithmetical progressions
uses even $O(\log|T|)$. Hence, if it is possible to ``balance''
any SLP up to $O(\log|T|)$ height, then the bound for working time
of our algorithm becomes $O(nm\log|T|)$.

It is interesting to consider more rules for generating texts,
since collage systems \cite{KiMaShTaShAr03} and LZ77 \cite{Zile77}
use concatenations and {\it truncations}. Indeed, as Rytter
\cite{Ry03} showed, we could leave only concatenations expanding
the archive just by factor $O(\log|T|)$. But there is a hope that
the presented technique will work directly for the system of
truncation/concatenation rules. More details will follow in a full
version.

We also claim that AP-table might be translated to a polynomial-sized SLP generating all occurrences of $P$ in $T$.

\section{New Algorithms for Related Problems}\label{other}

In this section we apply dynamic programming and our AP-table
routine for solving more classical problems: periods, covers and
fingerprinting. For highly compressible texts these algorithms
might be faster than ``unpack-and-solve'' approach. These results
show that pattern matching is not the only problem that we can
speed up. Hence, we get a more general understanding of how
processing on compressed texts really works. In this extended
abstract we present only the basic theoretical facts and short
sketches of the algorithms.

\subsection{Covers and Periods}\label{candp}

A {\it period} of string $T$ is a string $W$ (and also an integer
$|W|$) such that $T$ is a prefix of $W^k$ for some integer $k$. A
{\it cover} (originated from \cite{ApFaIl91}) of a string $T$ is a
string $C$ such that any character in $T$ is covered by some
occurrence of $C$ in $T$. Note that every cover/period is uniquely
determined by its length since by the definition they all are
prefixes of $T$. We use notation $t=|T|$ in this section.

\begin{center}
\includegraphics{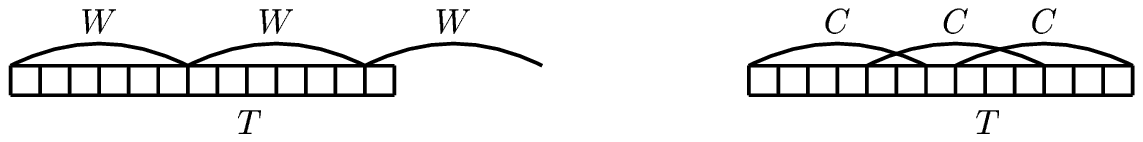}
\end{center}

Problem of {\it compressed periods/covers:} given a compressed string $T$, to find
a length of minimal period/cover and to compute a ``compressed''
representation of all periods/covers.

Theoretical facts serving as a basis for our algorithms:
\begin{enumerate}
  \item Given an SLP of size $n$ and for a given substring we can construct a linear-sized SLP
        (generating this substring) in linear time.
  \item Lengths of periods of the $t$-string between
        $t-\frac{t}{2^k}$ and $t-\frac{t}{2^{k+1}}$ form a single ar.pr.
  \item Any cover is a {\it border} (prefix and suffix of a given text). Any given text has a
        border of length $u$ iff it has a period of length $t-u$.
  \item If some border in the interval $[\frac{t}{2^{k+1}},\frac{t}{2^k}]$ is a cover, then
        all smaller borders in that interval
        are also covers.
\end{enumerate}

\begin{center}
\includegraphics{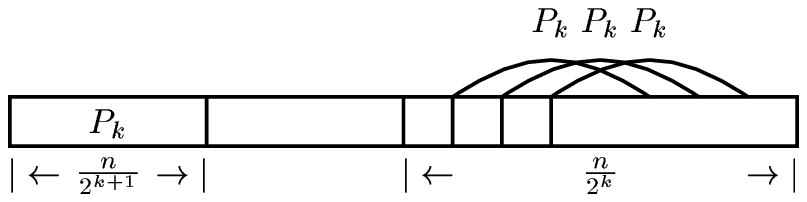}
\end{center}

{\bf Sketch of algorithm for compressed periods.} Using Fact 3 we
will search for borders instead of periods. We will separately
find all periods in each of $[\frac{t}{2^{k+1}},\frac{t}{2^k}]$
intervals. Let us fix some $k$. Step 1: we take a
$\frac{t}{2^{k+1}}$-prefix of $T$ (let us denote it by $P_k$) and
find all occurrences of this prefix in $\frac{t}{2^k}$-suffix of
$T$. By the Basic Lemma this is a single ar.pr. The distances
between starting positions of occurrences of $P_k$ and the end of
$T$ are {\it candidate} borders.

Step 2: we take a $\frac{t}{2^{k+1}}$-suffix of $T$ (let us denote
it by $S_k$) and find all occurrences of this suffix in
$\frac{t}{2^k}$-prefix of $T$. By the Basic Lemma this is a single
ar.pr. The distances between the start of $T$ and the ending positions of
occurrences of $S_k$ and the end of $T$ are {\it second candidate}
borders.

Fact: to obtain the set of lengths of all borders in our interval
we should simply intersect the sets of candidate and second
candidate borders. Let us estimate the complexity of processing
one interval. We apply FCPM algorithm to patterns that are substrings
of initial text $T$. By Fact 1 we can efficiently construct their SLP representation.
Moreover, the sizes of SLPs generating these patterns are larger than SLP generating $T$
by at most constant factor. Step 1 requires a single
FCPM call, step 2 also requires a single call, intersection of two
ar.pr. could be done in linear time. Hence, we need $O(n^3)$ time
for one interval and $O(n^3\log|T|)$ time for the whole compressed
periods problem.

\medskip
The compressed periods problem was introduced in 1996 in the extended abstract \cite{GaKaPlRy96}.
However, the full version of their algorithm (it works in $O(n^5\log^3|T|)$ time) was never published.
%Bad?

\medskip {\bf Sketch of algorithm for compressed covers.}
We will separately solve problem for cover size between
$\frac{t}{2^{k+1}}$ and $\frac{t}{2^k}$ for every $k$. As it was
shown in the previous algorithm, we can find all borders with
lengths in one such interval in time $O(n^3)$. Then using Fact 4 we
apply binary search with cover-check procedure.

Cover check procedure (for given compressed strings $C$ and $T$, to check whether $C$ is a cover of $T$):
\begin{enumerate}
  \item Compute AP-table for $C$ and $T$.
  \item For every intermediate text $T_j$ check by Local PM procedure
  whether all $|C|$-neighborhood of cut in $T_j$, excepting only characters that are less
  than $|C|$-close to the ends of $T_j$, are covered by $C$-occurrences.
  \item Check by Local PM procedure whether $|C|$-prefix and $|C|$-suffix of $T$ are
                completely covered by $C$-occurrences.
\end{enumerate}
We can prove by induction that $C$ is a cover iff we get only ``yes'' answers on Step 2 and Step 3.

Complexity analysis. Finding borders in all intervals requires $O(n^3\log|T|)$
time. Cover-check procedure requires $O(n^3)$ time, since Step 2
and Step 3 use only $O(n^2)$ time. We call cover-check procedure at most $\log |T|$ times in each
of $\log |T|$ intervals. Hence, the total complexity is $O(n^3\log^2|T|)$.

\subsection{Fingerprint table}\label{ft}

Let $\Sigma$ be an alphabet. A {\it fingerprint} is a set of used
characters of any substring in a text $T$. A {\it fingerprint
table} is a set of all fingerprints for a given text. The
algorithm for computing fingerprint table for usual (uncompressed)
strings is presented in \cite{AmApLaSa03}. Important fact: for any
string there might be at most $|\Sigma|+1$ (including $\empty$)
different fingerprints for all its suffixes (prefixes).

{\it Compressed fingerprint table:} given a compressed string $T$, to compute a fingerprint table.

\medskip {\bf Sketch of the algorithm.} For every $j$ compute all prefix-fingerprint, all suffix-fingerprint and
add to the table all fingerprints of ``cut-containing'' substrings
of $T_j$ by induction. At the end we clean the table from the
repeated fingerprints. Hence, we have $O(n|\Sigma|^2\log n)$
complexity and $O(n|\Sigma|^2)$ output size.
%Check!!!

\subsection{Hardness result}\label{hr}

{\it Compressed Hamming distance} (inequality version):
given integer $h$ and SLPs generating $T_1$ and $T_2$,
to check whether the Hamming distance (the number of characters which differ) between them is less than $h$.

\begin{theorem}
Compressed Hamming distance is \NP- and \coNP-hard (even inequality
version).
\end{theorem}
\begin{proof}
Recall a well-known \NP-complete \textsc{Subset Sum} problem \cite{GaJo79}:
Given integers $w_1,\dots,w_n,t$ in binary form, to determine
whether there exist $x_1,\dots,x_n \in\{0,1\}$ such that
\mbox{$\sum_{i=1}^n x_i\cdot w_i\quad =\quad t$}.

Let us fix some input values for \textsc{Subset Sum}.
We  now (efficiently) construct $P$, $Q$ and $h$ such that $HD(P,Q)<h$ iff
\textsc{Subset Sum} has a positive answer.

Let $s=w_1+\dots w_n$. Then we take
$$P=(0^{t}10^{s-t})^{2^n},\quad T=\prod_{x=0}^{2^n-1} (0^{\bar x\cdot \bar w}10^{s-\bar x\cdot \bar w}).$$
Here $\bar x \cdot \bar w = \sum x_iw_i$, and $\prod$ denotes
concatenation. The string $T$ (we call it {\it Lohrey string}) was
presented for the first time in Lohrey's work \cite{Lo04} and
then also used in \cite{LiLo06}. It was proved in \cite{Lo04} that
given input data of \textsc{Subset Sum} we can construct
polynomial-size SLP generating $P$ and $T$ in polynomial time.
But the inequality $HD(P,T)<2^{n+1}$ holds iff the \textsc{Subset Sum}
answer is ``yes''. Hence, the compressed Hamming distance is
\NP-hard. Inverting all bits in $P$ and taking $h=|P|-2^{n+1}+1$ we can prove \coNP-hardness.
\end{proof}

\section{Open Problems and Directions for Further Research}\label{conclusion}

\textbf{Algorithms.}
\begin{itemize}
 \item To speed up the presented algorithm for fully compressed pattern matching.
       {\it Conjecture:} $O(nm\log|T|)$ time is enough. More precisely, to show that computing new element of
       AP-table could be done in time $O(\log|T|)$.
 \item To construct a faster algorithm and/or
       with less memory requirements for compressed window subsequence problem. This problem was solved
       in paper \cite{CiGuLiMa06} in $O(nk^2)$ time and space.
       {\it Conjecture:} $O(nk^2)$  time, but $O(nk)$ space.
 \item To construct $O(nm)$ algorithms for (weighted) edit distance,
       where $n$ is original length of $T_1$ and $m$ is the size of SLP generating
       $T_2$. This will lead to immediate speed up from any ``super-logarithmic'' compression, since only
       $O(\frac{n^2}{\log n})$ classical algorithm is known for this problem.
 \item To speed up the presented algorithm for compressed fingerprint table or find an $n$-SLP-generated text
       with $\Omega(n|\Sigma|^2)$ fingerprint table.
%Think on last!!!
\end{itemize}

\textbf{Complexity.}
\begin{itemize}
 \item The membership of compressed string in a language described by extended regular language
       is \NP-hard. On the other hand it is in \PSPACE.
       To find exact complexity of the problem (repeated from \cite{Ry00}).
 \item The fully compressed subsequence problem is $\Theta_2$-hard \cite{LiLo06}.
       On the other hand it is in \PSPACE. To find exact complexity of the problem.
 \item The compressed Hamming distance is \NP- and \coNP-hard.
       On the other hand it is in \PSPACE. To find exact complexity of the problem.
 \item Is compressed weighted edit distance \NP-hard for any system of weights?
       What is exact complexity of the problem?
\end{itemize}

%\textbf{Open Problems: More general compression.}
%\begin{itemize}
% \item Suppose some texts are generated by a program $P$ (of more general type than SLP)
%       that works in $O(2^n)$ time. Could we always check the equivalence of two generated
%       texts in polynomial space?
%\end{itemize}

Other directions for further research include considering more classical string problems, more powerful models of
string generation and experimental study of new algorithms on compressed texts. Also it is extremely interesting
to find effective techniques besides dynamic programming.

\bibliographystyle{IEEEtranS}
\bibliography{arxiv06-final%,IEEEabrv
}
\end{document}